\documentclass[10pt,letterpaper,twocolumn]{article}
\usepackage{ol}
\usepackage{amsmath,amsfonts,epsfig}

\newcommand{\beq}{\begin{equation}}
\newcommand{\eeq}{\end{equation}}
\newcommand{\beqa}{\begin{eqnarray}}
\newcommand{\eeqa}{\end{eqnarray}}

\def\pra#1{{ Phys.\ Rev. A\/} {\bf#1}}

\def\prl#1{{ Phys.\ Rev.\ Lett.} {\bf#1}}

\begin{document}

\twocolumn[

\title{Fast Light in Fully Coherent Gain Media}
\author{B.D. Clader, Q-Han Park, and J.H. Eberly}
\affiliation{Department of Physics and Astronomy, University of Rochester \\ Rochester, NY 14627 USA}

\begin{abstract}
We analyze the propagation of fast-light pulses through a finite-length resonant gain medium both analytically and numerically.  We find that intrinsic instabilities can be avoided in attaining a substantial peak advance with an ultra-short rather than a long or adiabatic probe.
\end{abstract}
]

\maketitle

Studies of electromagnetically induced transparency (EIT) \cite{bib.Harris} have suggested ways that a strong laser field can dress a dielectric medium such that a weak probe pulse injected in the dressed medium will exhibit non-standard velocity of propagation \cite{Grobe-etal}. Many observations of so-called slow light \cite{bib.slow} and fast light \cite{bib.fast} in such dressed media have been reported.  Here the terms fast and slow refer to advanced or retarded propagation of a pulse peak, giving the appearance of pulse propagation faster or slower (even much slower) than the group velocity in the same medium without the benefit of dressing. 

In the case of fast light, which is our principal concern here, an effective technique \cite{bib.fast} is the manipulation, by strong coupling fields, of the spectral domain between two resonance lines by external {\em cw} fields to obtain very high dispersion in a very narrow spectral window through which a probe pulse will propagate with strongly altered velocity and anomalously low loss. The narrow spectral window can be obtained in more than one way \cite{bib.bigelow} but it is a universal feature of recent work and it severely limits the bandwidth of the probe pulse. Noise-related limitations on fast light have also been discussed \cite{bib.fundquestions}. We have considered modifications that may permit narrow-line limitations to be overcome, and we report results of a theoretical study of fast light with very short rather than very long probes. Fast light propagation of short pulses appears feasible with the use of appropriately configured transient gain media, given sufficient attention to natural instabilities.  

Fast light is associated with negative group velocities (for reviews and accounts of early experiments see Refs. \citeonline{bib.kryukov, bib.milonniB}).  Before 1990 theoretical examinations of this effect in gain media were largely semi-quantitative or numerical (e.g., see \citeonline{bib.basov, bib.icsevgi}), but propagation studies of superfluorescence by inverse scattering methods gave a number of valuable analytic insights  \cite{Gabitov-etal}. Here we have used an approach that provides analytic expressions for fast probe pulses and is flexible enough to treat different situations of potential experimental interest, including segmented and stacked dielectric media. Our solutions to the coupled nonlinear equations governing near-resonant pulse propagation are valid when probe pulses and transit times are short enough to allow upper-level relaxation effects to be ignored.  As we mentioned above, this is outside the zone of EIT, which uses long pulses and rapid upper level relaxation to produce a quasi-steady state, but we will show that the pulse advance associated with fast light can still be produced. 

\begin{figure}[h]
\begin{center}
\leavevmode
\includegraphics[height=0.9in]{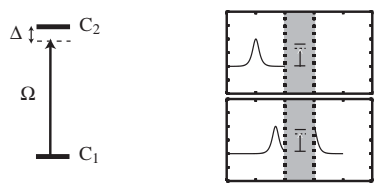}
\end{center}
\caption{\label{tla.fig} Two-level atom with level $1$ connected to level $2$ via the Rabi frequency $\Omega$ of a laser field detuned from resonance by an amount $\Delta$, and sketch of significant pulse advance in an inverted two-level gain medium.}
\end{figure}

To be as simple as possible, we model the dressed medium by a collection of inverted two-level atoms, similar to an approach taken by earlier workers \cite{bib.basov, bib.icsevgi}. A typical  two-level atom is shown in Fig. \ref{tla.fig}, and we consider media nominally subject to inhomogeneous broadening. The Hamiltonian of such a system in the rotating wave picture is given by
\begin{equation}
\label{eq.ham}
H = -\hbar\frac{\Omega}{2}|1\rangle\langle 2| -
\hbar\frac{\Omega}{2}|2\rangle\langle 1| + \hbar\Delta|2\rangle\langle 2|
\end{equation}
where $\Omega=2\mu E/\hbar$ is the Rabi frequency associated with the $1 \to 2$ transition and for simplicity it is assumed to be real, $\mu$ is the dipole matrix element, $E$ is the electric field envelope function, and $\Delta = \omega_{21} - \omega$ is the detuning of the laser below resonance.

We assume an atomic state of the form $|\psi (x,t)\rangle = c_1(x,t)|1\rangle + c_2(x,t)|2\rangle$ with $c_1$ and $c_2$ being the corresponding probability amplitudes.  Schr\"odinger's equation along with the Hamiltonian in Eq. \eqref{eq.ham} then gives
\begin{subequations}
\label{eq.schr}
\begin{align}
\frac{\partial c_1}{\partial t} & = i\frac{\Omega}{2} c_2 \label{schr1}
\\
\frac{\partial c_2}{\partial t} & = i\frac{\Omega}{2} c_1 - i\Delta
c_2 .\label{schr2}
\end{align}
\end{subequations}
Maxwell's equation written in the slowly varying envelope approximation (SVEA) yields an equation for the pulse envelope via the Rabi frequency:
\begin{align}
\label{eq.maxwell}
\left(c\frac{\partial}{\partial x} + \frac{\partial}{\partial t}\right)\Omega & = -ig \langle c_1c_2^* \rangle \\ \nonumber
& = -ig\int_{-\infty}^{\infty}d\Delta F(\Delta)c_1c_2^*
\end{align}
where $F(\Delta)=\frac{T_2^*}{\sqrt{2\pi}}e^{-(T_2^*\Delta)^2/2}$,
and $T_2^*$ is the inhomogeneous lifetime. The product $\pi gF(0)/c$ is the inhomogeneously broadened Beers absorption coefficient, and $g = N\mu^2\omega/\epsilon_0\hbar$, $N$ is the density of resonant atoms and $\omega$ is the probe pulse frequency.

The key differences to EIT theory are evident here -- there are no empirical relaxation terms included in the equations above, and there is no coupling laser to a third level. The first is justified by our desire to work with very short pulses, so we are thinking of pulses shorter than upper level dephasing times and lifetimes of say 10 ns. The second is based on the assumption that the conceptually simplest dressing of the medium is partial or complete initial inversion, which we will see to be sufficient for our purpose. We will see that the consequences of the differences are substantial.

We focus here specifically on stimulated rather than spontaneous emission, so all interactions are taken as induced, assuming negligible influence from spontaneous emission on the time scales of interest, including superfluorescence \cite{MB_superfluorescence}, which merits a specific qualification, mentioned later.  

We solve Eqns. \eqref{eq.schr} and \eqref{eq.maxwell} for a medium located between the finite positions $x_0$ and $x_1$.  The method we use, a variant of the B\"acklund transformation technique developed by Park and Shin \cite{bib.park}, allows us to find solutions for media that have a variable density of atoms. We have taken the simplest special case, constant $N$ inside the medium and $N=0$ outside, as is relevant to a thin inverted medium.  

Before the probe pulse arrives, the state of the atoms in the medium is $|\psi(x,t=-\infty )\rangle = |2\rangle$.  The exact SVEA pulse solutions are given by the following segmented expression that describes a fully continuous pulse in time and space:
\begin{equation}
\label{eq.pulsesol}
\Omega(x,t)=
\begin{cases}
\frac{2}{\tau}\textnormal{sech}\frac{1}{\tau}\left(t-\frac{x}{c}\right)
& \text{$x < x_0$}
\\
\frac{2}{\tau}\textnormal{sech}\frac{1}{\tau}\left(t-\frac{x}{v_g}+\phi_0\right)
& \text{$x_0 \le x \le x_1$}
\\
\frac{2}{\tau}\textnormal{sech}\frac{1}{\tau}\left(t-\frac{x}{c}+\phi_1\right)
& \text{$x > x_1$},
\end{cases}
\end{equation}
where $\tau$ is the nominal probe peak width. The group velocity $v_g$ is given by
\begin{equation}
\label{eq.groupvel}
1- \frac{c}{v_g} = \frac{g}{2}\int_{-\infty}^{\infty} F(\Delta) \frac{d\Delta}{\Delta^2+(1/\tau)^2} \to \frac{g\tau^2}{2},
\end{equation} 
where the final limit applies when $T_2^* \gg \tau$. In general, $v_g > c$ ~or ~$v_g < 0$, depending on $T_2^*$, the width $\tau$ of the probe peak and $g$ the atom-field coupling parameter.  The expressions $\phi_0 = -\tau^2gx_0/2c$ and $\phi_1 = \tau^2g(x_1-x_0)/2c$ are given in the long $T_2^*$ limit.  The corresponding probability amplitudes for the atoms in the medium are found to be given by $c_1(x,t) = i~\textnormal{sech}\phi(x,t)$ and $c_2(x,t) = - \textnormal{tanh}\phi(x,t)$, where $\phi(x,t) \equiv 
\frac{1}{\tau}\left(t-\frac{x}{v_g}+\phi_0\right)$.

\begin{figure}[h]
\begin{center}
\leavevmode
\includegraphics[height=2.2in]{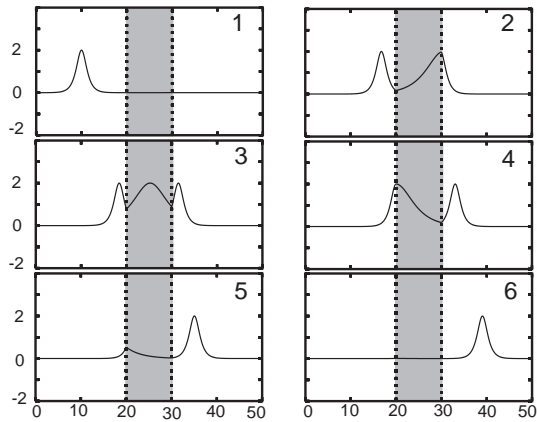}
\end{center}
\caption{\label{an_pulse.fig} Snapshots of the analytic solutions given in Eq. \eqref{eq.pulsesol}.  The horizontal axis is $x$ in units of $c\tau$.  The shaded region indicates the location of the medium, whose features are described in the text. For an EIT comparison, see Fig. 15 in Ref. \citeonline{bib.milonniP}.}
\end{figure}

Pulse and amplitude solutions for a fully coherent finite length medium were previously obtained by Andreev \cite{andreev} in the absence of Doppler broadening \cite{Rahman-Eberly98}. An adiabatic EIT-based example has been given by Milonni \cite{bib.milonniP}. Snapshots of our solution in Eq. \eqref{eq.pulsesol} are plotted in Fig. \ref{an_pulse.fig}, where one can clearly see that the pulse entering the medium is advanced -- in the third frame an outgoing peak leaves the medium before the input peak enters it.  In terms of initial peak widths this peak advance is slightly more than 10 peak widths.  In this particular case, the sample was chosen to be Rb vapor, using the D$_2$ line with these parameters: $T_2^* = 0.733$ ns, $g = 266$ ns$^{-2}$ and $\tau = 0.1$ ns, all of which appear to be within experimental reach.  They give an inverse Beer's length of $\alpha = \sqrt{\pi/2}gT_2^*/c \approx 8.15$ cm$^{-1}$ and a negative group velocity in the medium of $v_g/c \approx -3.27$, and the medium was taken approximately 250 absorption lengths deep.

The analytic pulse solutions (\ref{eq.pulsesol}) have $2\pi$ Area, and thus after interacting with the medium the pulse returns the atoms to their excited state.  This is a fully coherent but unstable situation, and the trigger for instability can come from several mechanisms. By assuming a very short pulse we have eliminated the main ones, which are associated with the medium's relaxation processes. However, we cannot realistically assume infinitely long pulse wings (as in the analytic sech functions), and the finite front edge of a pulse is an instability trigger that will always be present even if very weak (see \citeonline{Lamb, Gabitov-etal2} for a discussion of the consequences). 

To test sensitivity to this semiclassical trigger of instability we have generated numerical solutions to the same set of equations, with the same medium parameters. We used a short enough probe peak to justify ignoring significant medium relaxation, but we gave the pulse finite weak leading and trailing edges. The results are shown in the following figure.

\begin{figure}[h]
\begin{center}
\leavevmode
\includegraphics[height=2.2in]{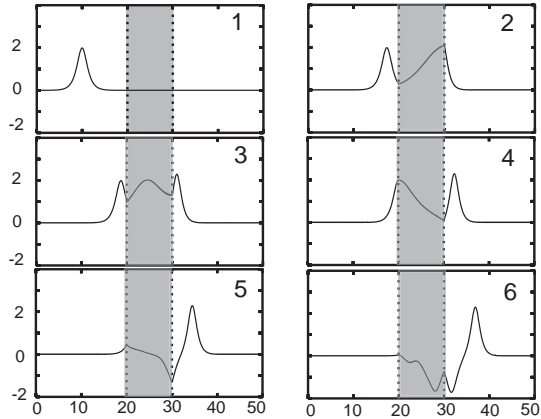}
\end{center}
\caption{\label{num_pulse.fig} Snapshots during evolution of the numerical pulse solution to Eqns. \eqref{eq.schr} and \eqref{eq.maxwell} with the same parameters as for Fig. \ref{an_pulse.fig}, but for a pulse initially entirely outside the medium with abrupt edges at $\pm$ 25 units from the peak.  The pulse advance is ~10 units forward, thus still well behind the front edge of the pulse.}
\end{figure}

In the snapshots of the numerical solution one again sees in frame 3 an advanced peak leaving the medium before the input peak enters, and there is a clear negative group velocity in the medium. The output pulse is relatively cleanly separated from background, and is essentially identical to that for the analytic solution, in both its extent of advance and in its Area, which is very close to $2\pi$.  The pulse ``ringing" trails the advanced pulse is radiation arising from the stimulated coherence imparted to the medium by the leading edge of the pulse.  The pulse must have Area = $\pm\pi$ to satisfy the two-level Area Theorem \cite{bib.mccall} for a gain medium, and numerical integration confirms that the total Area of the pulse changes from $2\pi$ to $\pi$ upon propagation through the inverted medium.  Burnham-Chiao ringing \cite{burnham-chiao} is not visible in Fig. \ref{num_pulse.fig}, but appears in later frames.

True spontaneous effects such as superfluorescence are excluded in any  treatment using classical fields but we have made a preliminary study of spontaneous effects on pulse advance by testing propagation in inverted media with small randomly distributed zero-average dipole moments, as a surrogate for the vacuum fluctuations that ``induce" spontaneous emission. These results indicate significant observable pulse advance (i.e., greater than 1 full input pulse width) even in the presence of such fluctuations.

In summary, we have examined a regime of semiclassical pulse propagation opposite that of EIT. To do this we chose pulse peak durations very short, and neglected medium relaxation times, so quasi-steady or adiabatic atomic states could not develop. A substantial peak advance is evident in our exact short-pulse solution, which has 2$\pi$ Area and infinitely extended tails on both sides of its brief intense peak. With inverted media, such exact solutions are well known to be unstable. The exact solution pulse served as a template against which our numerical propagation in equally inverted media, but with more realistic pulses with finite tails, could be compared. Our results indicate that a significant pulse advance (i.e., the fast light effect) appears reasonable to achieve, well-separated from the instability. In fact, Fig. \ref{num_pulse.fig} shows an equal advance in the numerical pulse as in the analytic result. 

It is interesting to note that since $\tau \sim T_2^*/7.33$, Doppler broadening played a limited role in this exercise. An increased inhomogeneous bandwidth would allow more rapid dephasing of atomic coherence and add a degree of  stabilization.

We wish to thank P.W. Milonni and I.R. Gabitov for useful discussions.  Permanent address of Q-Han Park is: Dept. of Physics, Korea University, Seoul, Korea. B.D. Clader acknowledges receipt of a Frank Horton Fellowship from the Laboratory for Laser Energetics, University of Rochester. Research supported by NSF Grant PHY 0456952.  The e-mail contact address is: dclader@pas.rochester.edu.

\end{document}